\documentclass[smallcondensed]{svjour3}     

\smartqed  

\usepackage{graphicx}
\usepackage{mathptmx}      
\usepackage{amsmath}

\newcommand{\fref}[1]{Fig.~\ref{#1}}
\newcommand{\Fref}[1]{Figure~\ref{#1}}
\newcommand{\sref}[1]{Sec.~\ref{#1}}

\newcommand{\bra}[1]{\langle#1|}
\newcommand{\ket}[1]{|#1\rangle}

\newcommand{\NOON}{\textrm{\sc noon}}
\newcommand{\JLM}{J_\textrm{LM}}
\newcommand{\JMR}{J_\textrm{MR}}

\newcommand{\phiL}{\phi_\textrm{L}}
\newcommand{\phiM}{\phi_\textrm{M}}
\newcommand{\phiR}{\phi_\textrm{R}}

\newcommand{\Sint}{S_\textrm{int}}
\newcommand{\Sdist}{S_\textrm{dist}}

\journalname{Few-Body Systems}

\begin{document}

\title{Entanglement in spatial adiabatic processes for interacting atoms}

\author{Albert Benseny \and Irina Reshodko \and Thomas Busch}

\institute{A.~Benseny \and I.~Reshodko \and Th.~Busch\at
              OIST Graduate University, 904-0495 Onna, Okinawa, Japan\\
              \email{albert.benseny-cases@oist.jp}       
}

\date{Received: date / Accepted: date}

\maketitle

\begin{abstract}
We study the dynamics of the non-classical correlations for few atom systems in the presence of strong interactions for a number of recently developed adiabatic state preparation protocols.
We show that entanglement can be created in a controlled fashion and can be attributed to two distinct sources, the atom-atom interaction and the distribution of atoms among different traps.
\keywords{Entanglement \and Spatial adiabatic passage}
\end{abstract}

\section{Introduction}

\label{intro}

Correlations between quantum particles are responsible for many properties of advanced solid state systems which cannot be derived from single particle behaviour \cite{Amico:08}. Recently these correlations were also recognised as a resource in quantum information and metrology, where they have to be deliberately created and engineered~\cite{Nielsen:2000}.  However, quantum correlations are known to be fragile, and preparing entangled states with high fidelity is often a difficult task~\cite{Monz:2011,Yao:12}.

The Hilbert space of interacting few or many particle systems becomes more complex and crowded, and preparing quantum states in such systems has an additional degree of complexity. However, over the last two decades, systems of ultracold atoms have been developed where control over almost all internal and external degrees of freedom is possible. By today, ground state systems of a handful of particles can be deterministically prepared in laboratories~\cite{Wenz:13}, and dynamical control of their center-of-mass degree of freedom is possible using time-dependent electromagnetic fields \cite{Murmann:15} .
The interaction between these ultracold atoms is usually short-range and for many species so-called Feshbach resonances exist, that can be used to adjust the interaction strength~\cite{Kokkelmans: 14}. 

One possible strategy for controlling interacting few atom systems is to generalise know single particle protocols. Out of these, one class that allows for high fidelities are adiabatic techniques, where the evolution follows one specific eigenstate at all times. For the control of the centre-of-mass of single particles, these are know as spatial adiabatic passage techniques~\cite{MenchonEnrich:16} and they have recently been shown to be suitable also for use for interacting systems~\cite{Benseny:16,Reshodko:17}. However, up to now no investigation has been undertaken that studies the dynamics of the non-local correlations during the adiabatic process and in this manuscript we will show how the behaviour of the entanglement can be understood for a few selected but representative protocols.

Our presentation is structured as follows.
In Sec.~\ref{sec:1DGases} we introduce the one-dimensional atomic system we study, and in \sref{sec:SAP} we present the spatial adiabatic passage processes we will consider.
In \sref{sec:nonclassic} we study the behaviour of the quantum correlations during these adiabatic processes, and conclude in \sref{sec:conc}.

\section{One-dimensional quantum gas model}
\label{sec:1DGases}
Recent experimental progress in cold atomic systems has allowed to create highly controllable few atom systems, that are effectively trapped in lower dimensions~\cite{Wenz:13}. One can therefore realise systems that correspond to model Hamiltonians that can be studied exactly with analytical as well as numerical tools. One example, and a good approximation to real experiments, are systems of $N$ interacting atoms trapped in a one dimensional geometry, where the interaction potential is assumed to be point-like. The Hamiltonian for such a system is 
\begin{equation}
	\hat H=\sum_{j=1}^N\left[-\frac{\hbar^2}{2m}\frac{\partial^2}{\partial x_j^2}+V(x_j,t)\right]+g\sum_{k>j}\delta(x_j-x_k),
\end{equation}
where $m$ is the mass of the atoms and the interaction strength is quantified by the coupling constant $g$, which can be related to the three-dimensional scattering length~\cite{Olshanii:98}. The position of each atom is described by $x_j$ and the time-dependent trapping potentials for the different spatial adiabatic passage protocols used below, $V(x,t)$, are given by piecewise harmonic oscillators of identical frequency $\omega$, that have a time-dependent position.
In what follows we will use natural units where $\hbar=m=\omega=1$.

The use of harmonic traps allows us to quantify the interaction strength in terms of the ground state energy of two atoms in a harmonic trap, $E_g$, through the relation~\cite{Busch:98}
\begin{equation}
	g=-\frac{2\sqrt{2}\Gamma(1-E_g/2)}{\Gamma((1-E_g)/2)}.
\end{equation}
In the non-interacting limit ($g=0$) this leads to $E_g=1$, corresponding to the ground state energy of two non-interacting atoms in a harmonic trap (twice the ground state energy of a single atom).
On the other hand, in the strongly interacting limit ($g\to\infty$, also known as the Tonks--Girardeau case), this gives $E_g=2$, which corresponds to the sum of energies of the ground and first excited states.
As a finite range of $E_g$ accounts for all repulsive interaction strenghts, we will plot our results below against $E_g$ rather than $g$.

In the following we will solve the time-dependent Schr\"odinger equation for the two-particle Hamiltonian by exact numerical integration, and quantify the entanglement inherent in the system by calculating the von-Neumann entropy given by~\cite{Nielsen:2000}
\begin{equation}
	S=-\sum_i \lambda_i\ln\lambda_i, 
\end{equation}
where the $\lambda_i$ are the eigenvalues of the reduced single particle density matrix
\begin{equation}
	\rho(x,x')=\int_{-\infty}^\infty \Psi^*(x,x_2)\Psi(x',x_2)dx_2,
\end{equation} 
and $\Psi(x_1,x_2)$ is the two-particle wavefunction at a fixed point in time.

\section{Spatial Adiabatic Passage}
\label{sec:SAP}

Spatial adiabatic passage (SAP) techniques~\cite{MenchonEnrich:16} are specific class of adiabatic methods that allow to manipulate the center-of-mass state of single particles in inhomogeneous potentials. They have been experimentally demonstrated recently~\cite{Taie:17}, making them an attractive quantum state engineering tool.
In their first incarnation~\cite{Eckert:04} they were developed as a direct translation of the well-known STIRAP technique in optics~\cite{Vitanov:17}.
However, atomic systems offer access to more degrees of freedom, which allowed to derive significant extensions, such as using systems of interacting particles.

In the following we will consider three different SAP protocols related to the transport of particles between traps, creation of spatial \NOON{} states, and splitting of clouds of atoms in a well defined way.
The basic idea behind SAP is to identify an eigenstate of the system that is a superposition of the given initial state and the desired final state and then change the relative weight of these two states by adjusting the external potential adiabatically. We will briefly review the required states in the following.

\subsection{Single particle SAP} 

To coherently transport a single atom from one trapping potential to another, SAP considers a setup consisting of three trapping potentials.
In our case, we will model them by harmonic oscillators of equal frequency (see \fref{fig:SAP+NOON}(a)).
Assuming that the atom is in the centre-of-mass ground state in the left trap, and that the evolution is adiabatic, one can reduce the system to considering only the lowest band, i.e. the three ground states in each trap, $\ket{\phiL}$, $\ket{\phiM}$, and $\ket{\phiR}$.
Using these states as a basis, the three-mode Hamiltonian can then be written as~\cite{MenchonEnrich:16}
\begin{equation}
\label{eq:H_SAP}
H_{3M} = -\frac{1}{2}
\begin{pmatrix}
	0      & \JLM{} & 0      \\
	\JLM{} & 0      & \JMR{} \\
	0      & \JMR{} & 0
\end{pmatrix},
\end{equation}
where the $J_{ij}$ describe the tunnel-couplings between neighbouring traps and the coupling between the outermost traps has been neglected as the tunnelling strength decays exponentially with distance. The ground state energies in each traps have been renormalised to zero. Transporting the atom to the trap on the right hand side can then be done by realising the existence of an eigenstate of the system of the form
	$\ket{D(\theta)} = \cos\theta \ket{\phiL} - \sin\theta \ket{\phiR}$,
which is the celebrated dark state. The mixing angle is given by $\tan\theta = \JLM{}/\JMR{}$, and one see that adjusting the individual tunnelling couplings can allow to switch a state from initially being located on the left hand side to finally being located on the right hand side by changing the mixing angle from zero to $\pi/2$. This leads to the counter-intuitive coupling sequence SAP and STIRAP are famous for, as this is achieved by moving the traps such that initially $\JLM{} \ll \JMR{}$, while at the end $\JLM{} \gg \JMR{}$.
In the same vein it is also possible to create a superposition state between the left and the right hand side trap, by adjusting the final mixing angle to be $\theta=\pi/4$, as $\ket{D(\pi/4)} = (\ket{\phiL} - \ket{\phiR})/\sqrt{2}$.
This can be achieved by symmetrically separating the traps, keeping $\JLM{}=\JMR{}$.

\subsection{Interacting particles transport and \NOON{} states} 

Due to the absence of strong decay channels in ultracold atomic lattice setups, the Hamiltonian in \eqref{eq:H_SAP} can be  generalised to interacting systems by making use of the possibility to create repulsively bound pairs~\cite{Winkler:06}.
In this case, $\{\ket{\phi_i}\}$ correspond to states where the two atoms are in the same trap, i.e.~$\ket{\phiL} = \ket{2~0~0}$, $\ket{\phiM} = \ket{0~2~0}$, $\ket{\phiR} = \ket{0~0~2}$, and the tunnelling couplings $J_{ij}$ will be given by their co-tunnelling amplitudes~\cite{Benseny:16,cotu4,cotu5,cotu6,cotu7,cotu8,cotu9}. 
However, the analogy between the two systems (single particle/interacting pair) requires an interaction strong enough such that the repulsively-bound pair is decoupled from states where the atoms are separated, but not so strong that they are coupled to higher excited states~\cite{Benseny:16}.

\begin{figure}
\centerline{\includegraphics[scale=0.8]{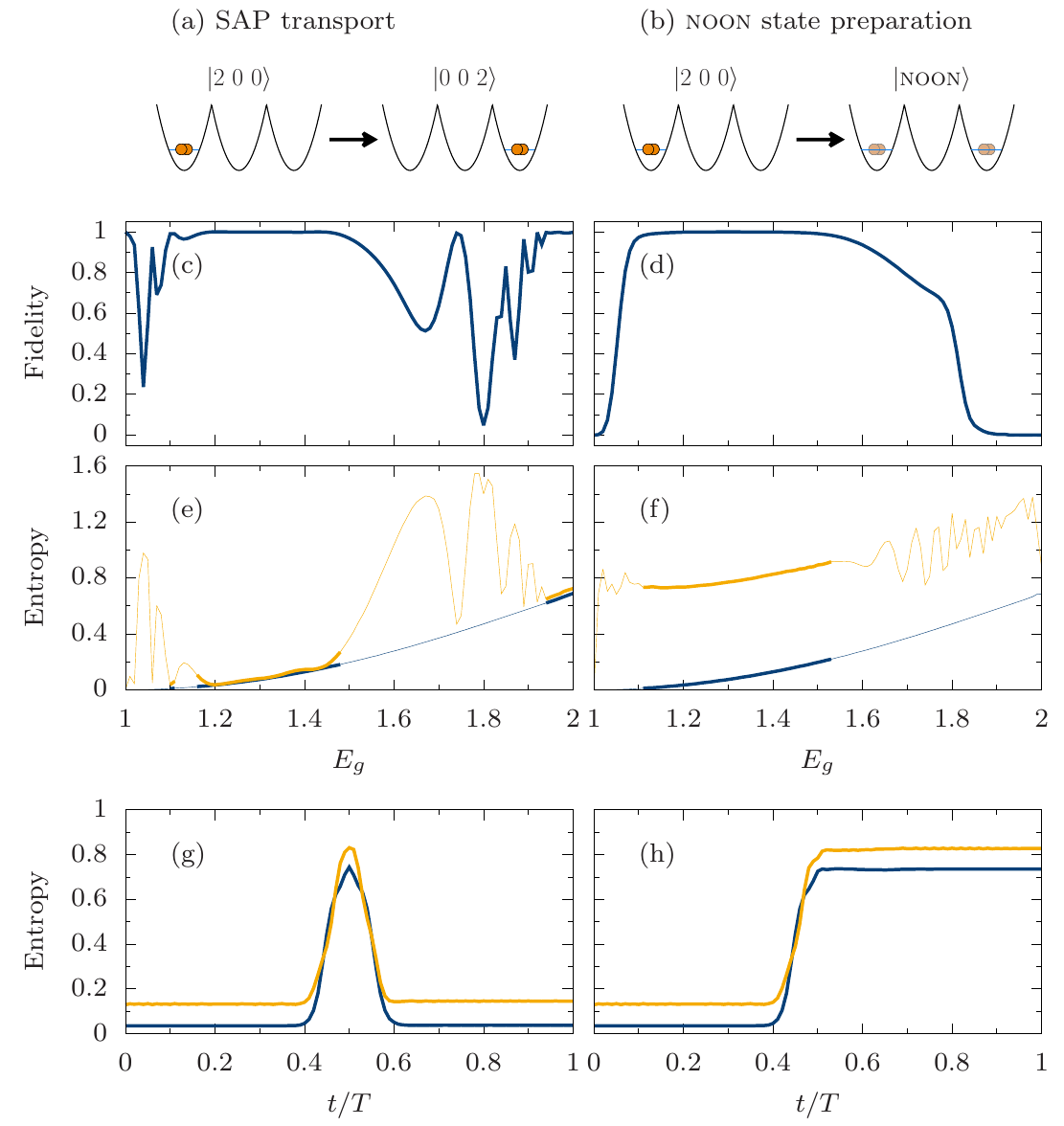}}
\caption{
(a) Sketch of the SAP transport process.
(b) Sketch of the \NOON{} state generation process.
(c-d) Fidelity of the transport/\NOON{} generation as a function of the interaction energy $E_g$.
(e-f) von Neumann entropy at the start (blue) and end (orange) of the transport/\NOON{} generation process as a function of $E_g$.
Thicker data points represent parameter values where the fidelity is above 98\%.
(g-h) Time evolution of the von Neumann entropy during  transport/\NOON{} generation for $E_g = 1.2$ (blue) and $E_g = 1.4$ (orange).
}
\label{fig:SAP+NOON}       
\end{figure}

This analogy allows to extend all single-particle SAP protocols to atom pairs, by performing the exact same trap movements as one would in the single atom case. For instance, the atom pair can be transported between the left and right traps by tuning $\theta$ between zero ($\ket{D(0)} = \ket{2~0~0}$) and $\pi/2$ ($\ket{D(\pi/2)} = \ket{0~0~2}$). More interesting, however, is the creation of a superposition by leaving the mixing angle at $\pi/4$, as this creates a \NOON{} state where the atom pair is in a superposition between the left and the right trap (see Fig.~\ref{fig:SAP+NOON}(b))
\begin{equation}
\label{eq:noon}
    \ket{D(\pi/4)} = \frac{1}{\sqrt{2}} (\ket{2~0~2} - \ket{0~0~2}) \equiv \ket{\NOON} .
\end{equation}
\NOON{} states are entangled states in which the macroscopic state of $N$ bosons is in a superposition of two distinct modes.
Such states have been generated for up to five photons~\cite{Afek:10} with promising applications in quantum technologies such as quantum sensing and quantum metrology~\cite{Lee:02}.
While generally states involving more than two particles are more valuable for different applications, the two-particle state we discuss is its smallest version of a \NOON{} state.
However, an extension of the above process to larger particle numbers are straightforward through higher-order co-tunnelling terms~\cite{Reshodko:17}.

\subsection{Particle separation}

While the above many-particle protocols are direct analogues of the single-particle settings, SAP techniques can also be used to split a cloud of interacting atoms and, in particular, separate a fixed number of particles from the initial state~\cite{Reshodko:17}.
This, however requires a slightly more complex setup, as redistributing the particles will lead to changes in the interaction energy which will affect the resonance conditions. To recover the resonance one can, for example, implement different onsite energy offsets for each trapping potential.

To determine these offsets, let us assume that we have initially $N$ atoms in the left trap, $\ket{N~0~0}$. This means that the two possible states after separating $n$ atoms in the other traps are given by $\ket{N-n~n~0}$ and $\ket{N-n~0~n}$, and  therefore  the energy shifts of the middle and right trap should be of the size of the interaction energy difference between having the $N$ atoms in the same trap or separated in $N-n$ and $n$ in different traps~\cite{Reshodko:17}.

After applying these two onsite shifts, the three states above therefore form a degenerate triplet, $\{\ket{\phi_i}\}$, and one can apply the SAP procedure to it. In this case, however, the tunnelling amplitudes are in general given by the $n$-particle co-tunnelling rates.
For this work, and without loss of generality, we start from $\ket{2~0~0}$, as it is sufficient to demonstrate the different processes that are involved in the entanglement dynamics. Since only a single particle should move from the left to the right trap, the same positioning sequence for the atom transport can be used to drive the system adiabatically into $\ket{1~0~1}$.

\section{Non-classical correlations}
\label{sec:nonclassic}

In the following we will examine the development of the entanglement inherent in the system during the processes outlined above. While one might naively think that the entropy, and therefore the entanglement, has to be conserved in an adiabatic process, this is not true during a dynamics where access to a number of degenerate states become available. In fact, we will show that the entanglement present stems from two different origins, with one being the inter-particle interaction and the other the distribution of the particles between the traps.

\subsection{Transport}

The first case we study is SAP transport for a pair of particles, sketched in \fref{fig:SAP+NOON}(a).
With the two particles initially in the left trap we simulate the SAP trap movement system and calculate the fidelity of the process, i.e., the occupation of state $\ket{0~0~2}$.
The results are shown in \fref{fig:SAP+NOON}(c) as a function of the interaction strength.
Since the dark state only exists in the absence of band crossings~\cite{Benseny:16}, the process is only successful over a finite range of interaction energies between about 1.1 and 1.5.
Outside of this range of interactions, level crossings appear which decrease the process fidelity.
The only exceptions are the non-interacting case and the Tonks--Girardeau limit, where
in the non-interacting case the transport of the two atoms becomes independent. 
In the Tonks--Girardeau limit, the interacting bosons can be mapped onto non-interacting fermions, and the SAP process can be decomposed into two independent processes, one for each single particle. 

The values of the von Neumann entropy before and after the transport process are shown in \fref{fig:SAP+NOON}(e), over the full range of interaction strengths.
We are only interested in the entropy for system parameters where the transport fidelity is perfect, i.e. the process is fully adiabatic, which corresponds to the values represented by a thicker curve.
One can see that the entropy increases with increasing interaction~\cite{Murphy:07}, however, when the transport process works, it has the same value at the beginning and end of the process.
This can be easily understood by realising that the initial and the final state are mirror images of each other, $\ket{2~0~0}$ and $\ket{0~0~2}$.

However, looking at the time evolution of the entropy during the SAP transport process, Fig.~\ref{fig:SAP+NOON}(g), one can see that the entropy actually increases and decreases during the process, at the time when all three trapping potentials are strongly coupled.
While this might at first be surprising, it is easy to understand this increase half-way through the process, as the system is in a superposition between the degenerate initial and the final state.

In fact, the maximal increase is found to be exactly $\ln 2$ when the superposition is completely symmetric.
This corresponds to the moment where the mixing angle is $\pi/4$, resulting in an instantaneous \NOON{} state.
If we were to separate the traps symmetrically at that point, the \NOON{} superposition would remain and the process would end with the atoms in this entangled state.
This is the next case we study.

\subsection{\NOON{} state preparation}

The creation of a \NOON{} state is sketched in \fref{fig:SAP+NOON}(b), and the fidelity of the process for different interactions is shown in \fref{fig:SAP+NOON}(d).
One can see from Figs.~\ref{fig:SAP+NOON}(c) and (d)  that the \NOON{} state process achieves high fidelities in the same interaction range as the transport,
i.e., for those interaction strengths which allow to avoid level crossings~\cite{Benseny:16}.
This is unsurprising, as the \NOON{} state generation is fundamentally half a transport process.
However, as the final state is now a superposition between two traps (the left most and the right most ones), the final entropy, and therefore entanglement, increases as compared to the initial value, see \fref{fig:SAP+NOON}(f) and (h).
It is interesting to note that the increase is independent of the interaction strength, and only depends on the distribution of the particles between the different traps.
In fact, the final density matrix is made from an equal superposition of states in the left and right trap and the entropy from this property is therefore
	$\Sdist=-2\left(\tfrac{1}{2}\ln\tfrac{1}{2}\right) \simeq 0.69$ .

It is also important to note that the \NOON{} preparation process does not succeed in either the non-interacting case or the Tonks--Girardeau limit.
This is because, as in the transport case, no entanglement between the two particles can be generated, as the dynamics of the two atoms become independent.
While this is trivial to see in the non-interacting case, the same happens in the Tonks--Girardeau limit due to the mapping of the atom dynamics to that of non-interacting fermions.

\begin{figure}
\centerline{\includegraphics[scale=0.8]{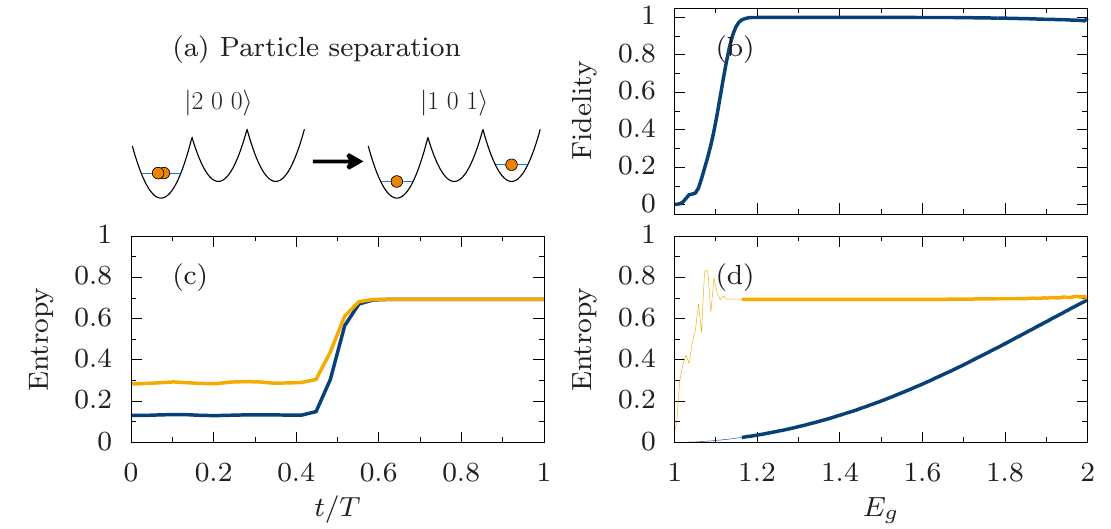}}
\caption{
(a) Sketch of the particle separation process.
(b) Fidelity of the scheme as a function of the interaction energy $E_g$.
(c) Time evolution of the von Neumann entropy during the particle separation process for $E_g = 1.4$ (blue) and $E_g = 1.6$ (orange).
(d) von Neumann entropy at the start (blue) and end (orange) of the separation process as a function of $E_g$.
Thicker data points represent those values where the fidelity is above 98\%.
}
\label{fig:SvN_saperition}
\end{figure}

\subsection{Particle separation}

The final SAP process we study is the separation of a particle pair between different sites~\cite{Reshodko:17}, sketched in 
\fref{fig:SvN_saperition}(a).
The fidelity of the process, shown in \fref{fig:SvN_saperition}(b), is very large as long as the interaction is above a certain threshold to ensure a band separation between the dark state and other bands~\cite{Reshodko:17}.

The dynamics of the entropy during the particle separation process for $E_g=1.4$ and $E_g=1.6$ is shown in \fref{fig:SvN_saperition}(c). 
The non-zero entropy of the initial states is due to the interaction-induced entanglement.
As the system follows the dark state, the entropy increases until it reaches $S(\ket{1~0~1})\approx0.69$, which corresponds to the entanglement purely due to the distribution of two indistinguishable particles between the wells.
Since the particles end up in different traps, the contribution of the interaction to the entanglement has become negligible \cite{Fogarty:11}, and in \Fref{fig:SvN_saperition} (d) we show that the final entropy of the $\ket{1~0~1}$ state is indeed independent of the interaction strength in the initial $\ket{2~0~0}$ state.
Note that the final value of the entropy is the same as the entropy increase in the case of the \NOON{} state preparation described above. 

\section{Conclusions}
\label{sec:conc}

We have shown that the dynamics of the non-classical correlations during a spatial adiabatic passage process can be simply understood by considering the two kinds of entropy: that due the interaction between the atoms and that due to the distribution of the atoms between the traps.
States where the two atoms are in the same trap, i.e., $\ket{2~0~0}$, have a von Neumann entropy which increases with the interaction $\Sint(E_g)$,  
ranging from zero in the non-interacting case (where the wavefunction is separable), to $\ln 2$ in the Tonks--Girardeau limit.
In this limit, due to the fermionisation of the atoms, states where the two atoms are in the same trap can be seen as them occupying the two lowest energy eigenstates.
The superposition due to the (anti)symmetrisation of the wavefunction, yields an entropy of $\ln 2$.

The other contribution to entropy is due to the distribution of the atoms between different traps.
For instance, states such as $\ket{101}$ are unaffected by the interaction, but have a constant von Neumann entropy of $\Sdist=\ln 2$ (see Appendix).
Moreover, \NOON{} states, such as \eqref{eq:noon}, will have both contributions, as the atoms occupy states where they interact, while they are in a superposition occupying different sites, $S_{\NOON} = \Sint(E_g) + \Sdist$.

\begin{figure}
\centerline{\includegraphics[scale=0.8]{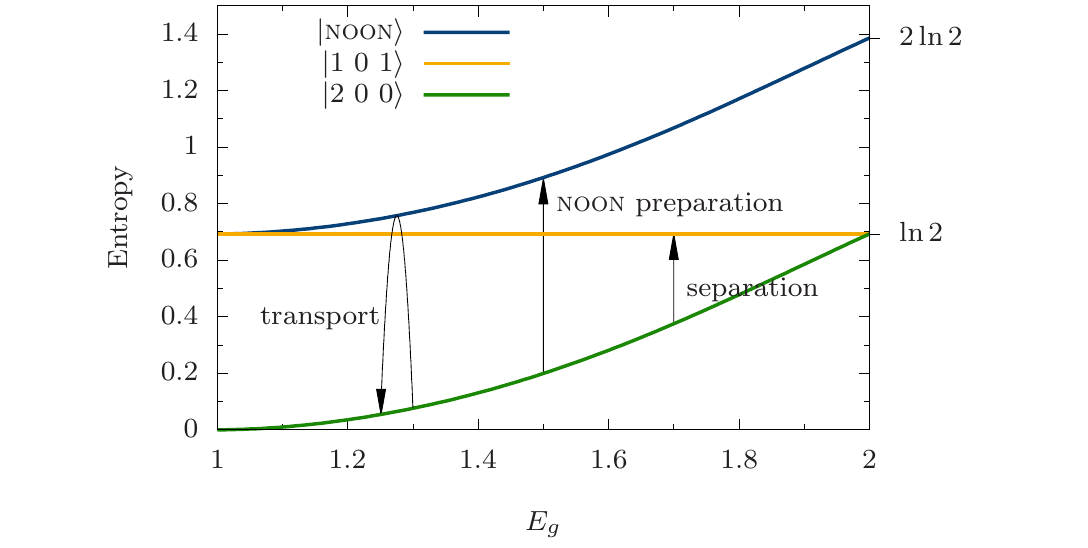}}
\caption{
Von Neumann entropy of states $\ket{200}$ (green), $\ket{101}$ (orange), and $\ket{\NOON}$ (blue) as a function of the interaction energy $E_g$.
The arrows represent the  changes in entropy during the spatial adiabatic passage processes studied in \sref{sec:nonclassic}.
}
\label{fig:states_SvN}
\end{figure}

A summary of the effects of the SAP processes we have studied on the von Neumann entropy is shown in \fref{fig:states_SvN}.
All processes start in state $\ket{2~0~0}$, and thus with entropy $\Sint(E_g)$. The transport process creates intermittent entanglement before returning to its initial value, whereas the \NOON{} state creation process allows to lock this additional entanglement in. In the separation process, the interaction entropy goes to zero while $\Sdist$ is created as the atoms appear in separate traps.

Although we have only studied two-particles processes above, the fundamental processes identified can straightforwardly be extended to systems with more particles (which, unfortunately, become numerically intractable).
For a splitting process with an initial cloud of $N$ particles in the left trap $\ket{N~0~0}$, a maximum value of the entropy around $\ln 2$ is achieved when the cloud is evenly split, $\ket{N/2~0~N/2}$, for $N$ even, or split into $\ket{\frac{N+1}{2}~0~\frac{N-1}{2}}$ for $N$ odd.
A \NOON{} state $(\ket{N~0~0}-\ket{0~0~N})/\sqrt{2}$, will also have an entropy of $\ln 2$.
If processes are designed that lead to a distribution of the atoms over more than two  traps, such as
$\ket{1~1~1~1~\ldots}$
or
$(\ket{N~0~0~0~\ldots} + \ket{0~N~0~0~\ldots} + \ket{0~0~N~0~\ldots}+\ldots)/\sqrt{N}$, the entropy would be given by
$\ln N$, and in the second case also the interaction entropy.

\appendix

\section*{Appendix: von Neumann entropy due to atom distribution}
\label{sec:VNentrCalc}

The von Neumann entropy associated with a state where atoms are in different traps can be calculated by writing the state in the atomic basis as
$\ket{101} = \frac{1}{\sqrt{2}}(\ket{L}_1\ket{R}_2+\ket{R}_1\ket{L}_2)$,
which results in a reduced density matrix for each atom
$\rho_1 = \frac{1}{2}(\ket{L}\bra{L}+\ket{R}\bra{R})$.
As $\rho_1$ is a diagonal matrix, the von Neumann entropy is given by
$S = -\tfrac{1}{2} \ln \tfrac{1}{2} -\tfrac{1}{2} \ln \tfrac{1}{2} = \ln 2 $.

Analogously, the \NOON{} state defined in \eqref{eq:noon},
$\ket{\NOON} = \frac{1}{\sqrt{2}}(\ket{L}_1\ket{L}_2 - \ket{R}_1\ket{R}_2)$,
results in the same reduced density matrix, and thus also has a von Neumann entropy of $\ln 2$.

\begin{acknowledgements}
This work was supported by the Okinawa Institute of Science and Technology Graduate University.
\end{acknowledgements}

\end{document}